\documentclass[english]{revtex4-1}
\usepackage[T1]{fontenc}
\usepackage[latin9]{inputenc}
\usepackage{geometry}
\usepackage{mathrsfs}
\usepackage{amsmath}
\usepackage{amssymb}

\makeatletter

\providecommand{\tabularnewline}{\\}

\usepackage{graphics}

\makeatother

\usepackage{babel}
\begin{document}
\title{A practicable guide to the quantum computation architectures}
\author{Hou Ian}
\address{Institute of Applied Physics and Materials Engineering, University
of Macau, Macau, China}
\author{Biao Chen}
\address{State Key Lab of IoT for Smart City, University of Macau, Macau, China}
\author{Wei Zhao}
\address{American University of Sharjah, Sharjah, UAE}
\begin{abstract}
The primordial model of quantum computation was introduced over thirty
years ago and the first quantum algorithms have appeared for over
twenty years. Yet the exact architectures for quantum computer seem
foreign to an undergraduate student major in computer science or engineering,
even though the mass media has helped popularize the terminologies
in the past decade. Despite being a cutting-edge technology from both
the theoretical and the experimental perspectives, quantum computation
is indeed imminent and it would be helpful to give the undergraduate
students at least a skeleton understanding of what a quantum computer
stands for. Since instruction-set architectures originated from classical
computing models are familiar, we propose analogously a set of quantum
instructions, which can be composed to implement renowned quantum
algorithms. Albeit the similarity one can draw between classical and
quantum computer architectures, current quantum instructions are fundamentally
incommensurable from their classical counterparts because they lack
the innate capability to implement logical deductions and recursions.
We discuss this trait in length and illustrate why it is held responsible
that current quantum computers not be considered general computers.
\end{abstract}
\maketitle

\section{Introduction}

Though descended from quantum mechanics and computer science, quantum
computation remains a difficult subject for both quantum physicists
and computer scientists. A major part of the difficulty arises from
the underlying counter-intuitive mathematical structure of quantum
mechanics, which even physicists find hard to incorporate into the
body of computer logic. Existing concepts in classical computation
are most likely not extensible to quantum computation, making it too
precocious a subject for typical computer science textbooks

Nevertheless, the counter-intuitiveness is exactly what enables an
algorithm, based on quantum mechanical principles, to be radically
different from a traditional algorithm. Although the motivation for
introducing the ''quantum'' leap remains the same as every other new
algorithm (that is to reduce complexity and save the computation time),
its implementation finds no siblings in existing algorithms. The change
is fundamental: the data unit is migrated from bit to qubit.

Qubit, short for quantum bit, is entirely a quantum mechanical concept;
that is to say it cannot be simulated by a classical program. A student,
therefore, equipped with the knowledge of Boolean algebra and discrete
mathematics, should experience no problem writing a classical computer
program but would find this knowledge insufficient to let them appreciate
quantum computation. However, it is not necessary for a student to
major in theoretical physics before they can appreciate this new way
of computation. This paper aims to distill the difficult concepts
into easier ones and make quantum computation accessible to undergraduate
students. A programmer ignorant of CPU layout or semiconductor theory
has no problem coding a program. Likewise, a quantum computer programmer
ignorant of quantum physics should be capable of composing a quantum
algorithm.

To achieve this goal, we start with the explanation of qubit, and
then proceed to explain how logic operations can be performed on a
qubit in Sec.~\ref{sec:Models}. To better understand how quantum
algorithms can be carried out on the qubits, we extend the concept
of instruction set to quantum computation and break down the algorithms
into smaller pieces of instructions in Sec.~\ref{sec:algorithms}.
This partitioning into finite series instructions elucidates the inner
functioning mechanism of each algorithm and allows one to see more
clearly how one algorithm is correlated to another when some of the
quantum instructions are shared by both algorithms.

In fact, IBM has released a set of assembly code instructions that
are readily executable on their soon-to-be commercially available
quantum computers, which is termed ''Open Quantum Assembly Language
(QASM)''~\citep{cross17}. QASM comprises a set of instructions more
fundamental than our approach here: with only a few exceptions, each
instruction corresponds conceptually to a quantum logic gate, which
is explained in Sec.~\ref{subsec:QLC_model}. The latter, in turn,
corresponds directly to a hardware implementation such that QASM can
be used to manipulate a qubit on a low level, very much like manipulating
the voltage level on a circuit node among the interconnected transistors
in a classical computer. Our approach is more pedagogic, aiming to
assist the understanding of existing quantum algorithms. In some sense,
each instruction covered in Sec.\ref{subsec:instruction_set} is a
higher-level composition of QASM instructions but bears a more intuitive
meaning. It is hoped that readers can recombine these instructions
to construct their own quantum algorithms.

Before discussing the quantum computation models, we give a brief
development history of quantum computation in Sec.~\ref{sec:history}.
After presenting the algorithms, we end the paper with a critique
on the current architectures of quantum computation in Sec.~\ref{sec:Critique},
discussing why they cannot yet be considered architectures for general
computation.

\section{Brief history\label{sec:history}}

The concept of the ''quantum computer'' was first introduced when
mathematicians tried to blend the tenets of quantum mechanics with
prototypical Turing machines in the 1980's~\citep{deutsch85,bennett89,yao93},
when physical systems based on quantum mechanical principles became
not merely measurable but, to a large degree, controllable. It was
already known by then that certain decision problems based on a probabilistic
Turing machine, i.e. allowing a margin of error probability, could
reduce algorithmic complexities.

Taking advantage of the probabilistic nature of quantum theory and
founded on the primordial concept of quantum Turing machines~\citep{yao93},
researchers have devised quantum versions of renowned algorithms in
the 1990's, such as large number factoring by Shor~\citep{shor97}
and search by Grover~\citep{grover96,grover97}. These algorithms
are founded on a set of quantum logic gates, which are inherited from
the classical concepts of logic gates, and are proven invariably faster
than their classical approaches. The parallelism permitted by quantum
states would sometimes cut the execution time on an exponential scale.
Deutsch has conceived an heuristic algorithm~\citep{deutsch92} to
illustrate the room of time saving, which is impossible for classical
machines.

The new millennium witnessed the substantiation of these ideas on
paper. Theoretical physicists proposed the conceptual \emph{qubit}
on experimentally realizable two-level atoms or spin-$\frac{1}{2}$
systems, on which logical operations can be realized as the interaction
of the qubit with a quantum field. Such systems constitute the quantum
logic computing (QLC) model, which we study below in Sec.~\ref{subsec:QLC_model}.
There are many candidate systems existing and the three most prominent
ones are trapped ions, nuclear spins in NV centers, and superconducting
qubits because of their high degree of manipulability. So far, the
superconducting qubit~\citep{orlando99,mooij99} is the most popular
candidate among academia and industries because this solid-state system
can be controlled and measured by programmable microwave equipment.
With the advent of circuit quantum electrodynamics~\citep{wallraff04},
quantum logic gates can be operated on these superconducting qubits
and Shor's algorithm was realized experimentally~\citep{lucero12}.

One of the disadvantages of the QLC model is its reliance on the manipulation
of all the qubits individually, making logical operations expensive
and scalability a nightmare. Therefore, some scientists took another
approach that was coined adiabatic quantum computing (AQC), to evade
the problem, which we study below in Sec.~\ref{subsec:AQC_model}.
Although the AQC model is still implemented on superconducting qubits~\citep{johnson11},
the computation method is distinct: the algorithmic result is obtained
through the minimization of system energy across an array of qubits.
Hence, the success of computation does not rely on the state of a
specific qubit. The drawback is that the adiabatic algorithm is heuristic
with no guarantee of efficiency improvement.

\section{Quantum computational models~\label{sec:Models}}

\subsection{Quantum logic computing (QLC) model~\label{subsec:QLC_model}}

\subsubsection{Qubit}

In a classical computer, the basic data unit is the bit (binary iteration).
It exists in the hardware as a circuit node, bearing either a low
or a high voltage level. It appears to the programmer as either the
integer 0 or 1 in the base-2 number system. It conveniently represents
the storage space of either the ``true'' or the ``false'' value of
predicate calculus such that the assertion of logical operations is
then possible to be carried out in a circuit.

A quantum system does not bear such a notion. Rather, the simplest
construct under the framework of quantum physics is a \emph{qubit}
(quantum bit). The construct partially retains its nomenclature of
``bit'' because it looks somewhat like a bit, to the extent that the
state of a qubit, usually denoted by the symbol $\left|\psi\right\rangle $,
can retain the state $\left|0\right\rangle $ or the state $\left|1\right\rangle $,
i.e. $\left|\psi\right\rangle =\left|0\right\rangle $ or $\left|\psi\right\rangle =\left|1\right\rangle $.
Nevertheless, $\left|0\right\rangle $ and $\left|1\right\rangle $
are just two extremal states permitted to be taken by the qubit and
not the only possible ones, unlike the classical bit. Instead, any
linear combination (as known as superposition) in the form of

\begin{equation}
\left|\psi\right\rangle =a_{0}\left|0\right\rangle +a_{1}\left|1\right\rangle \label{eq:superposition}
\end{equation}
is permissible as a state of the qubit $\left|\psi\right\rangle $.
In rigorous terminology of mathematics, $\left|\psi\right\rangle $
is regarded as a vector of a Hilbert space $\mathcal{H}$ while $a_{0}$
and $a_{1}$ are complex numbers which obey the normalization condition
$|a_{0}|^{2}+|a_{1}|^{2}=1$. Accordingly, $\left|0\right\rangle $
and $\left|1\right\rangle $ are a pair of orthogonal eigenvectors.

The orthogonality means that $\left|0\right\rangle $ and $\left|1\right\rangle $
has an inner product of zero, similar to the orthogonal vectors in
Euclidean spaces under the notion of dot product, and carries, to
some extent, the sense of logic as they seemingly designate opposite
extremes. The notion of eigenvectors is also similar to that of Euclidean
spaces. If we denote

\begin{equation}
\left|0\right\rangle =\left[\begin{array}{c}
1\\
0
\end{array}\right],\qquad\left|1\right\rangle =\left[\begin{array}{c}
0\\
1
\end{array}\right]
\end{equation}
as the basis vectors of a 2-dimensional space, they being eigenvectors
mean that there exist $2\times2$ matrices whose eigenvalues can be
obtained when applied on by these two vectors. The physical interpretation
is that one of the corresponding eigenvalue defines the energy of
the state represented by that vector, whereas the coefficients $a_{0}$
and $a_{1}$ designates a probability amplitude according to the so-called
Copenhagen interpretation. In other words, the norm squared $|a_{0}|^{2}$
is the probability that tells how likely one is to find $\left|\psi\right\rangle =\left|0\right\rangle $.
Therefore, a qubit is inherently probabilistic: it is unable to be
set to either $\left|0\right\rangle $ or $\left|1\right\rangle $
in a deterministic sense.

Despite the fundamental difference between bit and qubit, scaling
either data unit to a register is permitted. But unlike a classical
register, which can be constructed from concatenation of bits, a quantum
register is constructed from tensor products. For instance, $n$ qubits
can be used to construct a linear combination of $2^{n}$ states using
the binary number designation, i.e.
\begin{equation}
\left|\psi\right\rangle =\left(a_{0,1}\left|0_{1}\right\rangle +a_{1,1}\left|1_{1}\right\rangle \right)\otimes\left(a_{0,2}\left|0_{2}\right\rangle +a_{1,2}\left|1_{2}\right\rangle \right)\otimes\cdots=\sum_{i=0}^{2^{n}-1}a_{i}\left|i\right\rangle 
\end{equation}
where the normalization of all $a_{i}$ is still observed. The quantum
register will become the basic data unit to be employed in all algorithms
discussed below.

\subsubsection{Entanglement}

Quantum mechanics permits a peculiar feature between two or more physical
systems: entanglement. When multiple physical systems are entangled
or, equivalently speaking, a system of multiple subsystems is in an
entangled state, these subsystems are no longer independent, but inseparable
from each other. In other words, if one has two quantum registers
entangled, the state assumed by one register will have strong implication
on the state obtained by the other register.

This special property of entanglement is endowed by the innate mathematical
structure that dictates the formulation of quantum mechanics. When
two registers are present in a quantum computing system, the way to
describe their states simultaneously is to pair up their individual
Hilbert-space vectors through a mathematical procedure called tensor
product. For example, if the first register assumes the state $\left|\alpha\right\rangle $
while the second the state $\left|\beta\right\rangle $, then their
joint state is written as $\left|\alpha\right\rangle \otimes\left|\beta\right\rangle $.
Since the two registers are part of one computing system, we can equivalently
express the state of the system to be $\left|\alpha,\beta\right\rangle $.
Except for the choice of notation, these two expressions describe
the exact same thing. We have seen above that quantum mechanics postulates
a linear superposition principle for quantum states. Suppose each
of the two registers is one qubit, which individually can assume either
the $\left|0\right\rangle $ state or the $\left|1\right\rangle $
state. Then, it is possible to form the joint state 
\begin{equation}
\left|\psi\right\rangle =\frac{1}{\sqrt{2}}\left|0,1\right\rangle +\frac{1}{\sqrt{2}}\left|1,0\right\rangle \label{eq:ent_state}
\end{equation}
for the two quantum registers combined. This particular state, known
as the Bell state, is an entangled state and implies that whenever
the first register assumes the state $\left|0\right\rangle $, the
second register the state $\left|1\right\rangle $; whenever the first
register assumes the state $\left|1\right\rangle $, the second register
the state $\left|0\right\rangle $. Notice that this state can no
longer be separated into a tensor product, meaning we can no longer
determine the linear combination of each register alone.

This entanglement property plays a crucial role in developing quantum
algorithms. As we shall see in later sections, each of the quantum
algorithms as known today rely more or less on the formation of entangled
states between two or more registers. The importance and usefulness
of entanglement in terms of computation can be understood from an
analogy with associative memory. Like the example state given above,
once two registers are entangled and thus inseparable, the states
of one register would develop a definitive (not necessarily one-to-one)
correspondence with the states of the other, i.e. 0 to 1 and 1 to
0. In general, the correspondence can be illustrated by the Table~\ref{tab:entanglement}.
\begin{table}
\begin{tabular}{|c||c|c|c|c|c|c|c|c|}
\hline 
$\left|R1\right\rangle $ & $\left|0\right\rangle $ & $\left|2\right\rangle $ & $\left|7\right\rangle $ & $\left|3\right\rangle $ & $\left|5\right\rangle $ & $\left|6\right\rangle $ & $\left|4\right\rangle $ & $\left|1\right\rangle $\tabularnewline
\hline 
\hline 
$\left|R2\right\rangle $ & $\left|c\right\rangle $ & $\left|a\right\rangle $ & $\left|d\right\rangle $ & $\left|d\right\rangle $ & $\left|e\right\rangle $ & $\left|f\right\rangle $ & $\left|b\right\rangle $ & $\left|a\right\rangle $\tabularnewline
\hline 
\end{tabular}

\caption{Entanglement of one quantum register $\left|R1\right\rangle $ with
another register $\left|R2\right\rangle $.\label{tab:entanglement}}
\end{table}

The first row shows the eigenstates that comprise the superposed state
for the first register, while the second shows that of the second
register. Given the correspondence, we can immediately see that if
the first register is in, say, state $\left|5\right\rangle $, the
second register would definitely be in state $\left|e\right\rangle $.
Conversely, if the second register is in, say, state $\left|d\right\rangle $,
the first register would definitely adopt either state $\left|3\right\rangle $
or $\left|7\right\rangle $. Therefore, the joint states for a quantum
computing system will act like an associative memory, in which if
the data in one register is known, the data in the other register
is instantaneously given. In classical computers, the association
between two variables is universally useful since many functions demanded
in a computer can be reduced to such association between values stored
across multiple registers. For instance, multiplication is nothing
but associating a value called ``product'' with one value called ``multiplicand''
and with another value called ``multiplier.'' In other words, given
a multiplicand and a multiplier, the goal of multiplication is to
find or search for a value that associates with these two numbers.
A search in a database is even more straight-forward, given the association.
In the ``quantum'' associative memory illustrated in the Table~\ref{tab:entanglement},
the entangled states ($\left|0\right\rangle $ to $\left|7\right\rangle $
for $\left|R1\right\rangle $ and $\left|a\right\rangle $ to $\left|f\right\rangle $
for $\left|R2\right\rangle $) are given in a disarrayed manner. This
disorder is to illustrate that, unlike classical computers, quantum
computers do not necessarily have to adopt an order for addressing
the data in a memory system. The linear superposition principle holds
the implication that all states are born ordinally equal (neither
$\left|0\right\rangle $ has to be the first state nor $\left|7\right\rangle $
the last), though cardinally unequal (at any given moment, some state
has a higher probability while some other has a lower probability).
The uniformity in ordinality is, as we shall see in the algorithms
below, what grants the parallelism of and simplify the data processing
in a quantum computer.

\subsubsection{Quantum logic gates}

In a classical computer, the execution of an algorithm is carried
out through a series of logical operations: NOT, AND, OR, etc. On
the level of assembly code, each operation corresponds to a machine
instruction. On the hardware level, each operation corresponds to
a logical gate with input and output signal wires such that an algorithm
can be conceptually converted into a multi-stage gate diagram. Since
the input and output wires naturally correspond to hardware circuit
wire nodes, the gate diagram decides on the circuit wiring necessary
to carry out an algorithm execution.

Analogously, when quantum computation was first devised, researchers
had in mind a series of quantum logical gates that would be connected
to constitute a complete quantum algorithm~\citep{deutsch89,NIELSEN}.
Such gates include the Controlled-NOT or CNOT gate, Hadamard gate,
Pauli gates, Toffoli gate, etc. Though it is tempting to consider
these gates as quantum counterparts of classical logic gates, the
analogy only remains on a superficial level. Quantum algorithms are
indeed decomposable into quantum logic gates and these gates can be
graphically represented as diagrams with quantum states as inputs
and outputs. Nevertheless, unlike classical logic gates, which pertain
to the Boolean algebra as the foundation for carrying out the predicate
calculus, quantum logic gates have no such algebraic foundation and
cannot be configured into gates that carry logical meanings on a human
intentional level. Rather, all quantum logic gates are essentially
transformation matrices, which are square unitary matrices applicable
on quantum states when quantum states are regarded as vectors.

For example, the NOT gate carries the logical meaning of negation
for a given statement or predicate. If the truth value of that statement
is stored in a Boolean variable, it flips the value. There is, however,
no such negating gate for quantum states. Even though one can find
a matrix that transforms the state $\left|0\right\rangle $ to $\left|1\right\rangle $
(this matrix is called Pauli-$X$ gate), the transformation loses
the meaning of negation when the matrix is operated on quantum states
such as Eqs.~\eqref{eq:superposition} and \eqref{eq:ent_state}.

The most conspicuous difference between the two types of gates are
their reversibility. Classical logic gates are irreversible, meaning
the number of outputs are always less than the number of inputs, e.g.
AND has one output and two inputs. One can only conduct the logical
deduction along the forward direction, but not in reverse. From an
information-theoretic perspective, classical logic operations are
energy-consuming and entropy-reducing. On the other hand, since quantum
logic gates are unitary matrices in essence, they are reversible as
there always exists an inverse matrix for a given unitary matrix.
No information would be lost in carrying out quantum logic operations
and the number of inputs and outputs match, e.g. CNOT gate has two
inputs and two outputs. The lack of correspondance between existing
quantum logic gates and an algebra for predicate calculus lies at
the heart of the difficulty in promoting quantum systems as general
computation systems. We will discuss this problem in more detail in
Sec.~\eqref{sec:Critique}.

\subsection{Adiabatic quantum computing (AQC) model~\label{subsec:AQC_model}}

In contrast to the QLC model, another model knowned as adiabatic quantum
computing (AQC)~\citep{farhi00} also employs qubit as the basic
unit but has no reliance on the quasi-logic operations explained above.
Rather, it takes advantage of a cluster of qubits collectively to
perform the desired computation and its name is derived from the quantum
adiabatic process. While the classical adiabatic process signifies
a thermodynamic process where the evolution of the system is not accompanied
by the absorption or release of heat energy (i.e. enthalpy), its quantum
counterpart underlines the variation of a quantum system where the
probability distributions (i.e. $a_{0}$ and $a_{1}$ of Eq.~\eqref{eq:superposition})
remains unchanged but the energy levels denoted by $\left|0\right\rangle $
and $\left|1\right\rangle $ and corresponding to distinct eigenvalues
of energies change. Such a variation is possible if the process is
undertaken sufficiently slow, approaching the so-called adiabatic
limit.

Therefore, given a cluster of qubits with any initial configuration,
which is usually characterized by a temperature parameter for the
energy levels of the individual qubits, the common temperature can
be tuned for all qubits simultaneously through an adiabatic process.
The final configuration denotes a ground state where the eigenvalue
of energy for the cluster as a whole is minimized. Since the approach
to the ground state is accompanied by a negative temperature gradiant,
this adiabatic process simulates annealing and mirrors a class of
optimization problems in computation known by the same name. That
means, at the end of a simulated annealing algorithm, the ground state
would be the resulting output, connoting an optimized configuration
of a physical system mappable to AQC, such as the optimizated interconnect
patterns of an integrated circuit. Because the time it takes to approach
the ground state depends on the particular initial configuration of
each problem concerned, the annealing algorithms are always heurisitic
and the complexity reduction through AQC cannot be determined \emph{a
priori} like the algorithms solvable by QLC.

\subsection{Quantum instruction set\label{subsec:instruction_set}}

To better understand either the QLC model or the AQC model, we introduce
in this section a set of seven instructions. These seven instructions
form a powerful quantum instruction set that is sufficient to perform
significant quantum algorithms. We decompose three famous quantum
algorithms in the next section into instructions from this set.

\subsubsection{INI $R$}
\begin{itemize}
\item Description: initialize $R$ in quantum state $\left|R\right\rangle $
to the zero state, i.e. let $\left|R\right\rangle \to\left|0\right\rangle $.
\end{itemize}

\subsubsection{QFT $R$}
\begin{itemize}
\item Description: apply a quantum Fourier transform on $R$ 
\item Given a state vector $\left|R\right\rangle =\sum_{j=0}^{N-1}a_{j}\left|j\right\rangle $
or in the matrix form
\begin{equation}
\left|R\right\rangle =\left[\begin{array}{c}
a_{0}\\
\vdots\\
a_{N-1}
\end{array}\right]
\end{equation}
the resulting vector after the transformation is 
\begin{equation}
\mathrm{QFT}(\left|R\right\rangle )=\sum_{k=0}^{N-1}b_{k}\left|k\right\rangle \qquad\qquad\mathrm{where}\;b_{k}=\frac{1}{\sqrt{N}}\sum_{j=0}^{N-1}a_{j}\exp\left\{ i2\pi\frac{jk}{N}\right\} 
\end{equation}
\item Example: assume we have a two-qubit register $\left|R\right\rangle =a_{0}\left|00\right\rangle +a_{1}\left|01\right\rangle +a_{2}\left|10\right\rangle +a_{3}\left|11\right\rangle $,
which can be written as $\left|R\right\rangle =\sum_{j=0}^{3}a_{j}\left|j\right\rangle $
if we write the binary numbers in digits; then $b_{k}=\sum_{j=0}^{3}a_{j}\exp\left\{ i2\pi jk/4\right\} $.
To be specific, 
\begin{align}
b_{0} & =\frac{1}{2}\left[a_{0}+a_{1}+a_{2}+a_{3}\right]\\
b_{1} & =\frac{1}{2}\left[a_{0}+a_{1}e^{i\pi/2}+a_{2}e^{i\pi}+a_{3}e^{3i\pi/2}\right]\\
b_{2} & =\frac{1}{2}\left[a_{0}+a_{1}e^{i\pi}+a_{2}e^{2i\pi}+a_{3}e^{3i\pi}\right]\\
b_{3} & =\frac{1}{2}\left[a_{0}+a_{1}e^{3i\pi/2}+a_{2}e^{3i\pi}+a_{3}e^{9i\pi/2}\right]
\end{align}
In the special case where $a_{0}=1$ and $a_{1}=a_{2}=a_{3}=0$, we
have $b_{0}=b_{1}=b_{2}=b_{3}=1/2$.
\end{itemize}

\subsubsection{REA $R$}
\begin{itemize}
\item Description: read the state of the register $R$.
\item The chance of finding the value $x$ stored in $R$ follows the probability
distribution of each state $\left|x\right\rangle $ of $\left|R\right\rangle $.
\item After reading, the state of $\left|R\right\rangle $ collapses into
$\left|x\right\rangle $, i.e. $\left|R\right\rangle \to\left|x\right\rangle $
where all other possible states vanish.
\end{itemize}

\subsubsection{ENT $R1,R2,M$}
\begin{itemize}
\item Description: entangle the register $R1$ to $R2$ using transform
mapping $M$.
\item After the operation, each state $\left|j\right\rangle $ of $R1$
is entangled with state $\left|M(j)\right\rangle $ of $R2$. The
state distribution of $R1$ is not changed and the state distribution
of $R2$ is calculated based upon its association with $R1$.
\item Example: consider the mapping of modular exponentiation $M(j)=a^{j}\mathrm{mod}N$.
If $N=39$ while $a=7$, then values of $M(j)$ are shown in table~\ref{tab:transform_map}.
\begin{table}
\begin{tabular}{|c||c|c|c|c|c|c|c|c|c|c|c|c|c|}
\hline 
$j$ & 0 & 1 & 2 & 3 & 4 & 5 & 6 & 7 & 8 & 9 & \multicolumn{1}{c|}{10} & 11 & 12\tabularnewline
\hline 
\hline 
$M(j)$ & 1 & 7 & 10 & 31 & 22 & 37 & 25 & 19 & 16 & 34 & 4 & 28 & 1\tabularnewline
\hline 
\end{tabular}

\caption{Mapping of the modular exponentiation from $j$ to $M(j)$.\label{tab:transform_map}}
\end{table}
\end{itemize}
i.e. $\left|R1,R2\right\rangle =a_{0}\left|0,1\right\rangle +a_{1}\left|1,7\right\rangle +a_{2}\left|2,10\right\rangle +\cdots$.

\subsubsection{DIF $R,N$}
\begin{itemize}
\item Description: given an integer $N$, diffuse the states with higher
(concentrated) probabilities in the register $R$ into the states
with lower (diluted) probabilities
\item The operation is equivalent to the transformation under the symmetric
matrix
\begin{equation}
\left[\begin{array}{cccc}
\frac{2}{N}-1 & \frac{2}{N} & \cdots & \frac{2}{N}\\
\frac{2}{N} & \frac{2}{N}-1 & \cdots & \frac{2}{N}\\
\vdots & \vdots & \ddots & \vdots\\
\frac{2}{N} & \frac{2}{N} & \cdots & \frac{2}{N}-1
\end{array}\right].
\end{equation}
\end{itemize}

\subsubsection{PHA $R,\phi,n$}
\begin{itemize}
\item Description: perform a phase rotation of angle $\phi$ on $n$-th
state of the register $R$
\item The operation is equivalent to the transformation under the symmetric
matrix
\begin{equation}
\left[\begin{array}{ccc}
\begin{array}{ccc}
1 & 0 & 0\\
0 & 1 & 0\\
0 & 0 & \ddots
\end{array} & \cdots & \scalebox{3.5}{0}\\
\vdots & e^{i\phi} & \vdots\\
\scalebox{3.5}{0} & \cdots & \begin{array}{ccc}
\ddots & 0 & 0\\
0 & 1 & 0\\
0 & 0 & 1
\end{array}
\end{array}\right]
\end{equation}
\end{itemize}

\subsubsection{ANN $h,J$}
\begin{itemize}
\item Description: use quantum annealing to minimize the dimensionless energy
of the Ising model with parameter vector $h$ and parameter matrix
$J$:
\begin{equation}
\mathcal{E}(s|h,j)=\sum_{i\in V(G)}h_{i}s_{j}+\sum_{(i,j)\in E(G)}J_{i,j}s_{i}s_{j}
\end{equation}
where the quantum spin number $s_{i}$ is either $1$ or $-1$. $i$
indexes the vertices $V(G)$ of the associated graph $G$ fixed by
the device whereas $(i,j)$ indexes allowed pairwise interactions
given by edges $E(G)$ of this graph. Both $h_{i}$ and $J_{i,j}$
are real-valued dimensionless coefficients.
\end{itemize}

\section{Quantum algorithms\label{sec:algorithms}}

\subsection{Shor's integer factoring algorithm}

\subsubsection{Purpose of the algorithm}

Shor's approach to large integer factoring relies on the periodicity
(a.k.a. multiplicative order) of a random-base integer modulus of
the integer to be factored, i.e. if we denote the integer to be factored
by $N$ and pick a random base $x$, then we want to find the period
$r$ such that
\begin{equation}
x^{r}=1\mod N.
\end{equation}
Once $r$ is determined, $\gcd(x^{r/2}-1,N)$ will be a factor of
$N$ where the gcd can be found using an algorithm, such as Euclid's
method, that has much less complexity than those of the factoring
algorithms. To see why, consider that the congruence relation above
allows one to write $(x^{r/2}-1)(x^{r/2}+1)=mN$, where $m$ denotes
the quotient. Since $x^{r/2}-1$ and $x^{r/2}+1$ differ by 2 and
cannot be both factors of $m$, $x^{r/2}-1$ must contain at least
one factor of $N$ and this factor can be determined by the gcd. Hence,
the factoring problem is reduced to the problem of order finding.

Given $x$ and $N$, to find the order $r$, one needs to perform
the modular exponentiation $x^{j}(\mod N)$ until $j=r$ is found,
whose time complexity Shor showed can be reduced exponentially at
the expense of a logarithmic space complexity when the computation
is carried out on a quantum state. Shor showed that the maximum value
of $j-1$ needed to cover the testing range is $2^{n}$ where $N^{2}\leq2^{n}<2N^{2}$.
Therefore, two quantum registers $\left|R1\right\rangle $ and $\left|R2\right\rangle $
are required, where the first one storing the values of $j$ is entangled
with the second one carrying the modular exponential value $x^{j}(\mod N)$.
When this is done, the job of finding the correct $\left|j\right\rangle $
in $\left|R1\right\rangle $ for the period $r$ is accomplished by
applying a quantum Fourier transform on the joint state $\left|R1,R2\right\rangle $.

In other words, a quantum computer executing Shor's algorithm is actually
carrying out two operations: i) entanglement with modular exponentiation
and ii) quantum Fourier transform. We show the detailed steps below,
assuming $N=9$, $x=4$, and $j$ ranges from 0 to $2^{7}-1$ as an
example.

\subsubsection{Quantum assembly code}
\begin{enumerate}
\item INI $R1$
\begin{itemize}
\item Effect: $\left|R1\right\rangle =\left|0\right\rangle $.
\end{itemize}
\item INI $R2$
\begin{itemize}
\item Effect: $\left|R1\right\rangle =\left|0\right\rangle $.
\end{itemize}
\item QFT $R1$
\begin{itemize}
\item Effect: $\left|R1\right\rangle =\frac{1}{\sqrt{2^{7}}}\sum_{j=0}^{2^{7}-1}\left|j\right\rangle $.
\end{itemize}
\item ENT $R1,R2,x^{j}\mod N$
\begin{itemize}
\item Effect: $R1$ stays unchanged.
\item $R2$ is entangled to each state of $R1$ given by the map $f(j)=x^{j}\mod N$,
i.e.
\begin{equation}
\left|R1,R2\right\rangle =\frac{1}{\sqrt{2^{7}}}\sum_{j=0}^{2^{7}-1}\left|j,f(j)\right\rangle 
\end{equation}
where the pair $j$ and $f(j)$ are given by the table~\ref{tab:mod_exp}.
\begin{table}
\begin{tabular}{|c||c|c|c|c|c|c|c|c|c|c|c|}
\hline 
$j$ & 0 & 1 & 2 & 3 & 4 & 5 & 6 & 7 & 8 & $\cdots$ & 127\tabularnewline
\hline 
\hline 
$f(j)$ & 1 & 4 & 7 & 1 & 4 & 7 & 1 & 4 & 7 & $\cdots$ & 4\tabularnewline
\hline 
\end{tabular}

\caption{Transformation mapping from $j$ to $f(j)$ for entangling modular
exponentiated states.~\label{tab:mod_exp}}

\end{table}
\end{itemize}
\item REA $R2$
\begin{itemize}
\item Effect: $R2$ collapses into one $\left|f_{0}(j)\right\rangle $ state
among all $\{\left|f(j)\right\rangle \}$ states
\item $R1$ is reduced to a superposition state of all $\left|j\right\rangle $
eigenstates associated with $\left|f_{0}(j)\right\rangle $. Since
in the example assumed, $r=3$, we can write the superposition as
\begin{equation}
\left|R1\right\rangle =\frac{1}{\sqrt{43}}\sum_{m=0}^{42}\left|3m+c\right\rangle 
\end{equation}
given $c\in\{0,1,2\}$ as the residue in congruence with the period.
\end{itemize}
\item QFT $R1$
\begin{itemize}
\item Effect: the state $\left|R1\right\rangle $ of the first register
becomes
\begin{equation}
\left|R1\right\rangle =\frac{1}{\sqrt{43\times2^{7}}}\sum_{m=0}^{42}\sum_{k=0}^{2^{7}-1}\exp\left\{ i2\pi\frac{k(3m+c)}{2^{7}}\right\} \left|k\right\rangle .
\end{equation}
\item The probability of finding $\left|R1\right\rangle $ in one particular
$\left|k\right\rangle $ state among all $2^{7}$ states is
\begin{align}
\mathscr{P}(k) & =\frac{1}{43\times2^{7}}\left|\sum_{m=0}^{42}\exp\left\{ i2\pi\frac{k(3m+c)}{2^{7}}\right\} \right|^{2}\nonumber \\
 & =\frac{1}{43\times2^{7}}\left|\sum_{m=0}^{42}\exp\left\{ i2\pi\frac{3mk}{2^{7}}\right\} \right|^{2}\nonumber \\
 & =\frac{1}{43\times2^{7}}\left|\frac{\exp\left\{ i2\pi\frac{3\cdot43k}{2^{7}}\right\} }{1-\exp\left\{ i2\pi\frac{3k}{2^{7}}\right\} }\right|^{2}
\end{align}
where the second line is obtained because the imaginary $\exp\{i2\pi kc/2^{7}\}$
vanishes when the norm is taken and the third line completes the summation
over the geometric series.
\end{itemize}
\item REA $R1$
\begin{itemize}
\item Effect: $\left|R1\right\rangle $ collapses into one of the $\left|k\right\rangle $
state out of $2^{7}$ states
\item The likelihood of observing a particular $\left|k\right\rangle $
state is exactly the probability $\mathscr{P}(k)$, which maximizes
when $3k/2^{7}$ approaches unity.
\end{itemize}
\end{enumerate}
The last readout yields the desired result of $k$ where
\begin{equation}
k\approx\frac{2^{7}}{3}
\end{equation}
is about 43. Then the correct order $r=3$ can be inferred from the
$k$ value, i.e. $2^{7}/k\approx3$.

\subsection{Grover's search algorithm}

\subsubsection{Purpose of the algorithm}

Grover\textquoteright s algorithm is a search algorithm, which means
it gives the address in terms of a quantum ``index'' state from a
given quantum ``data'' state. The address is stored in the first quantum
register $\left|R1\right\rangle $ and the data in the second quantum
register $\left|R2\right\rangle $. The key is to correlate the address
with the data in one-one correspondence through entanglement. Then
when the data is fed, the system collapses onto the joint $\left|R1,R2\right\rangle $
state that corresponds to the data, giving the correct address in
$\left|R1\right\rangle $. The detailed process is the following.

\subsubsection{Quantum assembly code}
\begin{enumerate}
\item INI $R1$
\item INI $R2$
\begin{itemize}
\item Effect: to clear both registers to zero and prepare them for further
operations, similar to what occurred in Shor\textquoteright s algorithm
\end{itemize}
\item QFT $R1$
\begin{itemize}
\item Effect: to uniformly distribute probability in all possible indexing
values, from 1 to the maximal $N$, for the first register as an indexing
register.
\end{itemize}
\item ENT $R1$, $R2$, $M$
\begin{itemize}
\item Effect: to associate each indexing value in $R1$ to a unique value
to be searched in $R2$. In other words, states of $R1$ are analogous
to addresses while states of $R2$ are analogous to data.
\item The elements $M_{mn}$ of the transformation matrix $M$ is used to
specify the association.
\end{itemize}
\item for $i=1:\sqrt{N}$
\begin{enumerate}
\item PHA $R2$, $\pi$, $x$
\item DIF $R2$, $N$
\item Effect: through the iteration of this loop, the quantum state for
the value $x$ to be searched will have an iteratively increasing
probability weight while all other states will exhibit iteratively
decreasing probability weights. After $\sqrt{N}$ steps of iteration,
the correct quantum state will have a probability asymptotically close
to unity.
\end{enumerate}
\item REA $R1$
\begin{itemize}
\item Effect: one can read out the corresponding index state in $\left|R1\right\rangle $
that associates with the data state $\left|x\right\rangle $.
\end{itemize}
\end{enumerate}

\subsection{Deutsch-Jozsa's heuristic algorithm}

\subsubsection{Purpose of the algorithm}

Deutsch-Jozsa (DZ) algorithm is arguably the first quantum algorithm,
which predates Shor\textquoteright s algorithm and Grover\textquoteright s
algorithm. Though the algorithm does not carry an imminent practicality
for its function, it purposely illustrates the complexity advantage
of a quantum algorithm over its classical counterparts.

The DZ algorithm accomplishes this illustration through a deliberately
imposed mathematical function that transverses its entire input space
or domain. Its output space or codomain is the rather simple $\left\{ 0,1\right\} $
space. That is, for a natural number $N$, we assume the discrete
valued function $f(n):\mathbb{Z}_{2N}\to\mathbb{Z}_{2}$. The goal
is to determine which of the following two statement about $f$ is
true:

either (i) $f(n)\neq0$ and $f(n)\neq1$;

or (ii) $\left\{ f(0),f(1),\dots,f(2N-1)\right\} $ not contain exactly
$N$ zeros.\\
In short, the problem is to iterate through all $2N$ input values
and count the number of zeros obtained. No zeros, $N$ zeros, or $2N$
zeros constitute one case. All other scenarios constitute the complementary
case. Using two registers $\left|R1\right\rangle $ and $\left|R2\right\rangle $,
the quantum algorithm follows.

\subsubsection{Quantum assembly code}
\begin{enumerate}
\item INI $R1$
\item INI $R2$
\item QFT $R1$
\begin{itemize}
\item Effect: the three steps above are identical in purpose to those of
Shor's and Grover's algorithms.
\end{itemize}
\item ENT $R1$, $R2$, $M_{f}$
\begin{itemize}
\item Effect: the entanglement through transformation matrix $M_{f}$ fulfills
the mapping of $f$ in $\left|R2\right\rangle $ for each $i$, i.e.
$\left|i,j\right\rangle \to\left|i,j+f(i)\right\rangle $.
\item Since the second register $\left|R2\right\rangle =\left|0\right\rangle $
(i.e. $j=0$), the joint state of the two registers becomes
\begin{equation}
\left|R1,R2\right\rangle =\frac{1}{\sqrt{2N}}\sum_{i=0}^{2N-1}\left|i,f(i)\right\rangle .
\end{equation}
\end{itemize}
\item for $i=0:2N-1$
\begin{enumerate}
\item REA $R2$
\item PHA $R2,\pi f(i),f(i)$
\end{enumerate}
\begin{itemize}
\item Effect: read out the value of $f(i)$ for each $i$-th state in $R2$
and use the value to perform add a phase $e^{i\pi f(i)}$ on the state
$\left|f(i)\right\rangle $.
\item Since $e^{i\pi}=-1$, this loop essentially inverts the sign for each
state $\left|f(i)\right\rangle $ according to the value $f(i)$,
i.e.
\begin{equation}
\frac{1}{\sqrt{2N}}\sum_{i=0}^{2N-1}\left|i,f(i)\right\rangle \to\frac{1}{\sqrt{2N}}\sum_{i=0}^{2N-1}(-1)^{f(i)}\left|i,f(i)\right\rangle 
\end{equation}
\end{itemize}
\item ENT $R1$, $R2$, $M_{f}$
\begin{itemize}
\item Effect: Since $f(n)$ is a binary value function, one always has $f(i)+f(i)=0$.
Performing the entanglement operation again leads $\left|R2\right\rangle $
back to $\left|0\right\rangle $.
\item Essentially, we have
\begin{equation}
\left|R1,R2\right\rangle =\frac{1}{\sqrt{2N}}\sum_{i=0}^{2N-1}(-1)^{f(i)}\left|i,f(i)+f(i)\right\rangle =\frac{1}{\sqrt{2N}}\sum_{i=0}^{2N-1}(-1)^{f(i)}\left|i\right\rangle \otimes\left|0\right\rangle .
\end{equation}
\end{itemize}
\end{enumerate}
After executing the instructions above, the first register contains
all the information one needs to determine the truth value of statements
(i) and (ii). To see this, consider the inner product of $\left|R1\right\rangle $
with itself,
\begin{equation}
|\left\langle R1|R1\right\rangle |=\frac{1}{2N}\sum_{i=0}^{2N-1}(-1)^{f(i)},
\end{equation}
which is 0 when statement (i) is true and is 1 when statement (ii)
is true.

\section{Critique of current architectures~\label{sec:Critique}}

In the above sections, we have formally dissembled Shor's algorithm
into a set of high-level instructions, each of which can be further
reduced to a set of atomic instructions implementable on a superconducting-circuit
based quantum computer. It is not hard to see that even equipped with
these instructions, one cannot implement an arbitrary quantum algorithm
other than integer factoring at ease. This is because the given instruction
set lacks a sense of propositional logic implied in every classical
computer architecture.

Specifically, the sense of truth or falsity, which has always been
epitomically represented by 1 and 0 in Boolean logic, is not supplied
by quantum operations. The latter are really transformations of the
state vector representing a qubit and have no intrinsic logical implications.
Without ascertaining true or false as a terminal value in a conditional
statement, recursions cannot be implemented. Since recursive programming
is pivotal to classical computer programs, the lack of it deprives
the sense of automation when one talks about quantum computing. In
fact, many people have suspected that the quantum computer is at most
an ASIC-type device suitable only for a specific task, e.g. integer
factoring, and cannot be regarded as a general-purpose computer. The
reason behind can be attributed to this privation of logic and recursion.
Without the support of logic as the skeleton of quantum algorithms,
the specific algorithms explained in the sections above cannot be
dissembled into subroutines and then reassembled to become new algorithms
of the same class. The critiques here can be illustrated through a
comparison of the existing hardware for classical computers and quantum
computers.

\subsection{The classical computer}

All modern computers are based on devices fabricated on silicon. The
most basic device that can be fabricated is the PN-junction also known
as the diode (P and N, respectively, stands for the positively and
negatively doped silicon). It is, circuit-wise, a voltage-dependent
current source. Putting two such junctions back-to-back, one obtains
a sandwiched Source-Bulk-Drain structure. When a Gate-terminal over
an insulating oxide layer is deposited on top of this sandwiched structure,
one can control arbitrarily whether the current runs from Source to
Drain or not, thus a field-effect transistor is essentially accomplished.
The transistor is, thus, a voltage controlled switch and is the most
fundamental device of a classical computer.

When several such transistors are so wired that their Drains are commonly
connected to a suspended node, a binary memory device is essentially
accomplished. The Boolean logical 1 is obtained when an electric current
fills the suspended Drain node up to certain voltage prescribed by
the voltage-current relationship of the PN-junction between Bulk and
Drain. Adversely, the Boolean logical 0 is obtained when the current
reverses its direction to deplete the charges at Drain and sets its
voltage equal to ground.

Once the logical values 1 and 0 are set equivalent to physical high
and low voltages, elementary logical operations can be translated
into serial and parallel connections of multiple transistors together.
That is, logical gates are formed by feeding input signals (high or
low voltages) to the Gate terminals of these transistors to determine
whether currents would replete or deplete a drain node such that a
logical 1 (high voltage) or 0 (low voltage) would be obtained as the
output. For example, the classical logic NOT is translated into a
serial connection of two transistors where the input is connected
to the gates of both transistors whose sources are each in contact
with a high-voltage reservoir and the ground. The output is the terminal
of the two drains wired together.

If four transistors are given with two of the gates connected to one
input and the other two to another input, a NAND gate is made. Connecting
the NAND gate with the NOT above ($4+2=6$ transistors), one finds
the logical AND; connecting three NAND gates to two NOT gates ($12+4=16$
transistors), one finds the logical XOR (exclusive OR). In other words,
with a few transistors, we have implemented an adder with XOR giving
the sum bit and AND giving the carry-over bit of the result. Throughout
the computation, the logical values and, hence, the numeric values
at both the input and the output ends are designated by a set of high
and low voltages. The media that sustain the numeric information are
therefore the suspended circuit nodes that carry these electric potentials
and are not the transistors comprising the PN-junctions.

The wired circuits of the transistors make the instructions of a classical
computer well-defined, in the sense that each part of an instruction
corresponds uniquely to a specific part of a specifically wired circuit.
Consider, for example, a typical instruction: ADD InA, InB, Out. The
three letters ADD, signifying an addition, recall the specific adder
circuit consisting of 22 transistors, as explained above, for each
bit of InA, InB, and Out. Correspondingly, each bit of InA or InB
(Out) would become a specific node at the input (output) end of the
22-transistor circuit. The correspondence between the software and
the hardware is clear. But, as we will discuss below, the quantum
computer has not yet enjoyed this clear distinction, at least for
the implemented prototypes thus far.

The unique correspondence between a logical circuit and an instruction
not only facilitates the implementation of an extensive set of algebraic
instructions, but also makes logical determinations and, hence, branching
instructions for recursive programs possible. Consider, for instance,
a typical recursive function that ends its iteration when a variable
reaches a given integer value. If that variable is kept count in a
register $R1$, then the recursion loop always contains a statement
that calls the adder to increment $R1$ by one while keeping track
of its value to determine if the loop breaking condition is satisfied
during every iteration.

Therefore, since Kleene~\citep{kleene35} constructed the natural
number system on top of the first-order logic, every computer relies
on integers not only for direct calculation, but also for counting
the steps in executing algorithms. Offloading the job of counting
to a register is what endows the classical computer with the sense
of automation. In the picture of von-Neumann architecture, a classical
computer is a finite-state automaton that executes logical operations
and is equipped with a dedicated ``address'' register for keeping
track of the machine state.

\subsection{The quantum computer}

In contrast, the quantum computer, at its current state of development,
lacks such a device for keeping track of the machine state. As we
have dissembled the existing quantum algorithms in the sections above,
they are all equivalent to one-directional sequences of quantum state
transformations without any branching statements or recursions.

Therefore, they do not rely on an address register to keep track of
the quantum machine state and hence they do not yet possess the classical
sense of automata. One can sense that quantum devices which execute
these algorithms resemble a desktop calculator more than a programmable
computer, albeit one that performs state transformation rather than
arithmetic operations. It is for this reason that many computer scientists
do not regard quantum computers, architecturally speaking, as general-purpose
computers.

In terms of hardware, the current quantum computer design does not
correspond one-to-one to a classical computer. The most basic unit
of a quantum computer is, as already explained, a qubit. Unlike a
transistor, which is a gating or switching device, a qubit is rather
a storage or memory device. In other words, while we need several
transistors wired in a particular fashion to form a one-bit device
that can store one classical bit of information, the qubit is an indivisible
quantum device that can function as the memory unit for one qubit
of information . So far, memory is the only function a qubit can perform
and, further unlike transistors, qubits and even several qubits wired
together cannot function as quantum logical gates.

To carry out instructions or logical operations on the qubits, one
relies on external signals such as laser pulses or microwave pulses
to perform the relevant transformations. The sources of these external
signals, such as a laser diode or a microwave generator, can be regarded
as the controllers to the qubits. They are classical devices and serve
as the human-machine interfaces to current quantum computers. It is
these external sources which convert the quantum instructions depicted
in previous sections into specific sequences of pulses, which in turn
transform the qubit to a desired state.

We can therefore notice that:
\begin{enumerate}
\item programming is conducted on classical devices instead of the qubits;
\item the qubits do not actively perform the quantum instructions but are
passively operated by the instructions fed externally; and
\item though each qubit is described by a quantum state (i.e. the linear
superposition of $\left|0\right\rangle $ and $\left|1\right\rangle $),
the combined state of all the qubits in a quantum computer does not
constitute a machine state of an automaton.
\end{enumerate}
The reason for (iii) is that during the execution of an algorithm,
the states of the qubits are not read by the external sources to influence
the subsequent instructions to be fed. If the quantum computer possessed
a pointer similar to a classical computer, this quantum pointer would
only continue to the next instruction and never skip ahead or jump
backward.

Although we discussed quantum logic gates, such as Hadamard gates
and Toffoli gates, they cannot be seen as extensions of classical
logic gates since their compositions do not exhaust all the possible
logical combinations in the quantum state space while classical logic
gates (NAND or NOR, as explained above) do. More importantly, when
they are performed on qubits, the results are not true or false logic
termination values, but still states of qubits. Thus, programs, at
least in the classical sense, cannot rely on them to determine a branching
condition or the exiting condition of a recursion. The lack of a direct
mapping from superposition states in a Hilbert space into the ring
of integers or the Boolean algebra makes programmers who are accustomed
to coding classical algorithms unable to conceive of quantum algorithms
in a natural way.
\begin{acknowledgments}
H. I. thanks the support of FDCT of Macau under grant 065/2016/A2,
University of Macau under grant MYRG2018-00088-IAPME.
\end{acknowledgments}

\end{document}